\documentclass[12pt]{article}
\begin{document}


%
%

\title{Quantum Singularities in Spherically Symmetric, Conformally Static Spacetimes
}

\author{T. M. HELLIWELL \\ Physics Department, Harvey Mudd College\\
Claremont, California, 91711, USA \\ helliwell@hmc.edu
\and D. A. KONKOWSKI \\ Mathematics Department, U.S. Naval Academy, \\ 572C Holloway Road\\
Annapolis, Maryland, 21402, USA \\ dak@usna.edu}

\maketitle


\begin{abstract}
\noindent A definition of quantum singularity for the case of static spacetimes has recently been extended to conformally static spacetimes. Here the theory behind quantum singularities in conformally static spacetimes is reviewed, and then applied to a class of spherically symmetric, conformally static spacetimes, including as special cases those studied by Roberts,  by Fonarev, and by Husain, Martinez, and N\'u\~nez. We use solutions of the generally coupled, massless Klein-Gordon equation as test fields. In this way we find the ranges of metric parameters and coupling coefficients for which classical timelike singularities in these spacetimes are healed quantum mechanically.  

\end{abstract}


\section{Introduction}

Classical singularities, as characterized by the theorems of Hawking and Penrose, are ubiquitous in general relativistic spacetimes (see, e.g., \cite{HE}). The theorems do not necessarily indicate a divergence in the curvature, but rather geodesic incompleteness in otherwise maximal spacetimes. Spacetime geodesic incompleteness means, at least in the timelike and null cases, that classical particle paths come to an abrupt end. 

\par Classical singularities can be classified by their strengths \cite{ES,HE}. Quasiregular singularities are the mildest true singularity; they are topological in nature, basically holes in the fabric of spacetime. Conical singularities, as in idealized cosmic strings, are a good example. The other two types of singularities are stronger, curvature singularities. They are designated nonscalar or scalar depending on whether scalars in the curvature, such as the Ricci scalar and the Kretschmann scalar, diverge. Usually only $C^0$ scalars are considered, although there have been investigations of higher-order diverging scalar polynomial invariants (see, e.g., \cite{KH} and references therein). Nonscalar curvature singularities include those in whimper cosmologies and certain plane-wave spacetimes, whereas  scalar curvature singularities are the best-known, occurring at the center of black holes or the beginning of big bang cosmologies. Naked singularities, singularities not covered by an event horizon, of any of these types are even more troublesome; they are not only mathematical curiosities but have observable gravitational effects \cite{VNC,VE, VK}. 

\par The hope is that most singularities, especially naked singularities, can be ``resolved" or ``healed" in a complete quantum theory of gravity, and at least in the case of the two most prominent theories, string theory \cite{string-1, string} and loop quantum gravity \cite{loop}, there is a hint that this might be the case. For example, orbifolds are erased in string theory. What can we hope for in a complete or generic theory of quantum gravity? To answer this question we have stepped back and examined the simplest quantum generalization of classical singularities, the so-called quantum singularities.

\par Quantum singularities are generalizations of geodesic incompleteness to quantum wave packet ill-posedness. The idea originated with a paper by Wald \cite{wald} and was further developed by Horowitz and Marolf \cite{HM}. The basic idea is to study the behavior of wave packets in a spacetime, and see if they have a well-defined evolution, without placing boundary conditions at the location of the classical singularity that one hopes to heal quantum mechanically. As first proposed, the analysis was restricted to scalar (Klein-Gordon) wave packets in a static spacetime with a classical timelike singularity. That has since been extended to other fields (e.g., Maxwell and Dirac) \cite{HKA} and recently the authors have proposed an extension to conformally static spacetimes \cite{KH, KH-1}. Here we focus on the application to the conformally static case by studying a class of spherically symmetric spacetimes that includes as special cases those studied by Roberts \cite{R}, by Fonarev \cite{F, Maeda}, and by Husain, Martinez, and Nunez \cite{HMN}. 

\section{Quantum singularities in static spacetimes }

A static spacetime is quantum-mechanically (QM) \emph{non}singular if the evolution of a test scalar wave packet, representing the quantum particle, is uniquely determined by the initial wave packet, manifold and metric, without having to place boundary conditions at the singularity \cite{HM}. Technically, a static ST is QM-\emph{singular} if the spatial portion of the relevant wave operator, here the Klein-Gordon operator, is not essentially self-adjoint \cite{RS, RS-1} on $C_{0}^{\infty}(\Sigma)$ in the space of square-integrable functions $\mathcal{L}^2(\Sigma)$, where $\Sigma$ is a spatial slice.

\par A relativistic scalar particle with mass $M$ can be described by the positive-frequency solution \cite{HM} to the Klein-Gordon equation 

\begin{equation}
\frac{\partial^{2}\Psi}{\partial t^{2}} = - A\Psi
\end{equation}

\noindent in a static spacetime where the spatial operator $A$ is

\begin{equation}
A\equiv-KD^{i}(KD_{i})  + K^{2}M^{2}
\end{equation}

\noindent with $K=-\xi_\mu\xi^\mu$. Here $\xi^\mu$ is the timelike Killing field and $D_i$ is the spatial covariant derivative on a static slice $\Sigma$. The appropriate Hilbert space is $\mathcal{L}^{2}(\Sigma)$.  If we initially define the domain of $A$ to be $C_{0}^{\infty}(\Sigma)$, $A$ is a real, positive, symmetric operator and self-adjoint extensions always exist. If there is only a single, unique extension $A_E$, then $A$ is essentially self-adjoint. In this case, the Klein-Gordon equation for a free scalar particle takes the form

\begin{equation}
i \frac{d\Psi}{dt} =(A_{E})^{1/2}\Psi
\end{equation}

\noindent with 

\begin{equation}
\Psi(t) = \exp(-it(A_{E}^{1/2}))\Psi(0).
\end{equation}

\noindent These equations are ambiguous if $A$ is not essentially self-adjoint, in which case the future time development of the wave function is ambiguous. This fact led Horowitz and Marolf \cite{HM} to define quantum-mechanically singular spacetimes as those in which $A$ is not essentially self-adjoint.

\par Note that an operator $A$ is said to be self-adjoint if (i) $A = A^*$ and (ii) $Dom(A)$ = $Dom(A^*)$, where $A^*$ is the adjoint of $A$ and $Dom$ is short for domain \cite{RS, RS-1}. An operator is \emph{essentially} self-adjoint if (i) is met and (ii) can be met by expanding the domain of the operator $A$ or its adjoint $A^*$ so that it is true.

\par   One way to test for essential self-adjointness is to use the von Neumann criterion of deficiency indices \cite{VN, RS}, which involves studying solutions to the equation $A\Psi = \pm i\Psi$, where $A$ is the spatial portion of the Klein-Gordon operator, and finding the number of solutions that are square integrable ($i.e.$, $\in \mathcal{L}^2(\Sigma)$ on a spatial slice $\Sigma$) for each sign of $i$.  Another approach, which we have used before(see, e.g.,  \cite{ KH} and references therein) and will use here, has a more direct physical interpretation.  A theorem of Weyl \cite{RS, weyl} relates the essential self-adjointness of the Hamiltonian operator to the behavior of the ``potential" in an effective one-dimensional Schr\"odinger equation (made from the radial equation in a cylindrically or spherically symmetric spacetime), which in turn determines the behavior of the scalar-wave packet.  The effect is determined by a \emph{limit point-limit circle} criterion \cite{RS}.

The technique is straightforward for static spacetimes with timelike singularities. After separating the wave equation for a static metric, with changes in both dependent and independent variables, the radial equation can be written as a one-dimensional Schr\"odinger equation $Hu(x) = Eu(x)$  where the operator $H=-d^2/dx^2 + V(x)$ and $E$ is a constant, and any singularity is assumed to be at $x = 0$.  This form allows us to use the limit point-limit circle criteria described in Reed and Simon \cite{RS}.\\

\noindent $\mathbf{Definition}$.  \emph{The potential} $V(x)$  \emph{is in the limit circle case at}  $x = 0$ \emph{if for some, and therefore for all} $E$, \emph{all solutions of} 
$Hu(x) = Eu(x)$ \emph {are square integrable at zero.  If} $V(x)$ \emph{is not in the limit circle case, it is in the limit point case.}\\

\noindent A similar definition pertains for $x=\infty$: the potential $V(x)$ is in the limit circle case at $x=\infty$ if all solutions of $Hu(x)=Eu(x)$ are square integrable at infinity; otherwise $V(x)$ is in the limit point case at infinity.

\par There are of course two linearly independent solutions of the Schr\"odinger equation for given $E$. If $V(x)$ is in the limit circle case at zero, both solutions are square integrable ($\in \mathcal{L}^2(\Sigma)$)   at zero, so all linear combinations are square integrable $\in \mathcal{L}^2(\Sigma)$  as well.  We would therefore need a boundary condition at  $x=0$ to establish a unique solution.  If $V(x)$ is in the limit \emph{point} case, the  $\mathcal{L}^2(\Sigma)$    requirement eliminates one of the solutions, leaving a unique solution without the need of establishing a boundary condition at $x=0$.  This is the whole idea of testing for quantum singularities; there is no singularity if the solution in unique, as it is in the limit point case.  The critical theorem is due to Weyl,
\\[0.2cm]
\noindent $\mathbf{Theorem \ 1}$ (Theorem X.7 of Reed and Simon \cite{RS, weyl}).
Let $V(x)$ be a continuous real-valued function on $(0,\infty)$. Then  $H=-d^2/dx^2 + V(x)$ is essentially self-adjoint on $C_{0}^{\infty}(0,\infty)$ if and only if $V(x)$ is in the limit point case at both zero and infinity.
\\[0.2cm]
A useful theorem at infinity is 
\\[0.2cm]
\noindent $\mathbf{Theorem \ 2}$ (Theorem X.8 of Reed and Simon \cite{RS}).  \emph{Let $V(x)$ be a continuous real-valued function on $(0, \infty)$ and suppose that there exists a positive differentiable function $M(x)$ so that
\\[0.2cm]
(i) $V(x) \ge - M(x)$
\\[0.2cm]
(ii) $\int_1^{\infty} (M(x))^{-1/2} \ dx = \infty$
\\[0.2cm]
(iii) $M'(x)/(M(x))^{3/2}$ is bounded near $\infty$.
\\[0.2cm]
Then $V(x)$ is in the limit point case (complete) at $\infty$.}
\\[0.2cm]
\noindent A useful theorem near zero is
\\[0.2cm]
\noindent $\mathbf{Theorem \ 3}$ (Theorem X.10 of Reed and Simon \cite{RS}).  \emph{Let} $V(x)$ \emph{be continuous and positive near zero.  If} $V(x) \ge\frac{3}{4}x^{-2}$ \emph{near zero then} $V(x)$ \emph{is in the limit point case.  If for some} $\epsilon>0$, $V(x) \le(\frac{3}{4}-\epsilon)x^{-2}$ \emph{near zero, then} $V(x)$ \emph{is in the limit circle case.}\\

\noindent Theorem 3 states in effect that the potential is only limit point if it is sufficiently repulsive at the origin that one of the two solutions of the one-dimensional Schr\"odinger equation blows up so quickly that it fails to be square integrable. Another useful condition, as stated by Reed and Simon \cite{RS}, which does not require that $V(x)$ be positive, is that $- d^2/dx^2 + V(x)$ is limit circle at zero if $V(x)$ is decreasing as $x$ goes to zero.

Horowitz and Marolf used  the Hilbert space $\mathcal{L}^2$ when they studied the essential self-adjointness of the spatial Klein-Gordon operator in static spacetimes with classical timelike singularities \cite{HM}. Subsequently, Ishibashi and Hosoya  \cite{IH} used as the Hilbert space the 1st Sobolev space $H^1$; they then studied "wave regularity" of Klein-Gordon waves on static spacetimes with a classical timelike singularity. Here we follow the Horowitz and Marolf definition, as it uses the usual $\mathcal{L}^2$ Hilbert space of quantum mechanics.

\par By now many spacetimes have been tested to see whether or not quantum particles heal their classical singularities.   For example, we have studied quasiregular  \cite{KH3} and Levi-Civita spacetimes \cite{KHW, KRHW}, and used Maxwell and Dirac operators \cite{HKA} as well as the Klein-Gordon operator, showing that they give comparable results.  Cylindrically symmetric spacetimes were considered \cite{MM}, and Blau, Frank, and Weiss \cite{BFW} in addition to Helliwell and Konkowski \cite{HK4} have studied two-parameter geometries whose metric coefficients are power-laws in the radius  $r$  in the limit of small  $r$.  Pitelli and Letelier have considered the global monopole \cite{PL2}, spherical and cylindrical topological defects \cite{PL1}, BTZ spacetimes \cite{PL3}, and have recently extended their discussions along a new path to investigate cosmological spacetimes \cite{PL4, PL5}. They also have a review paper \cite{PL6}  on the mathematical techniques of quantum singularity analysis for static spacetimes along with numerous examples. Pitelli and  Saa \cite{PS} investigated quantum singularities in Horava-Lifshitz cosmology while Gurtug and co-workers examined the quantum singularity in Lovelock gravity \cite{MGH}, in a model of $f(R)$ gravity \cite{TG}, in an Einstein-Maxwell-Dilaton theory \cite{MHSG},  and  in a 2+1 dimensional magnetically charged solutions in Einstein-power-Maxwell theory \cite{MGHV}. Unver and Gurtug \cite{UG} studied quantum singularities in (2+1) dimensional matter coupled to black hole spacetimes. Seggev \cite{IS} studied possible extensions to stationary spacetimes. And, finally more recently Koehn \cite{Koehn} looked at relativistic wave packets in classically chaotic quantum billiards, a BKL-type scenario. A critical question in all of this work is: When is this use of quantum particles effective in healing classical singularities?

\section{Quantum singularities in conformally static spacetimes}

In a previous paper we described how to extend the methods for static spacetimes to the case of conformally static spacetimes.\cite{KH}  In particular, we considered conformally coupled scalar fields on conformally static spacetimes with timelike classical singularities.\footnote{Ishibashi and Hosoya \cite{IH} have already looked at such spacetimes using "wave regularity". We have explored \cite{KH, KH-1} a flat Friemann-Robertson-Walker spacetime containing an idealized cosmic string.} The natural form and separability of the wave equations on conformally static spacetimes was exploited. Here we elaborate on that extension.

\par A spacetime that admits a timelike conformal Killing vector field $W$ is known as conformally stationary 
\cite{constat}. As V. Perlick \cite{P} nicely summarizes,
\\[0.2cm]
\noindent ``If $W$ is complete and there are no closed timelike curves, the spacetime must be a product: $M\simeq R \times \hat{M}$, with a Hausdorff and paracompact 3 manifold $\hat{M}$ and $W$ parallel to the $R$ lines. If we denote the projection from $M$ to $R$ by $t$ and choose local coordinates $x = (x^{1}, x^{2}, x^{3})$ on $\hat{M}$, the metric takes the form

\begin{equation}
g = e^{2 f(t,x)} [(-dt + \hat{\phi}_{\mu}(x) dx^{\mu} )^{2} + \hat{g}_{\mu \nu}(x) dx^{\mu} dx^{\nu}]
\end{equation}

\noindent with $ \mu, \nu = 1, 2, 3. "$  
\\[0.2cm]
\noindent He goes on to define the more restrictive condition of  conformally static, 
\\[0.2cm]
\noindent ``If  $\hat{\phi} = \partial_{\mu}h$, where $h$ is a function of $x = (x^{1}, x^{2}, x^{3})$, we can change the time coordinate according to $t \mapsto t + h(x)$, thereby transforming $\hat{\phi}_{\mu}  dx^{\mu}$ to zero, i.e., making the surface $t$ = constant orthogonal to the $t$-lines. This is the conformally static case."  
\\[0.2cm]
Therefore, a conformally static spacetime $g_{\mu\nu} (x^\alpha )$ is related to a static spacetime $\gamma_{\mu\nu}( x^a)$ by a conformal transformation $C(\eta)$ of the metric.  Here $C(\eta)$ is the conformal factor, where $\eta$ is the conformal time, related to the time $t$ by $dt = C d\eta$.  Simply put,  $ g_{\mu\nu}(x^\alpha) = C^2(\eta) \gamma_{\mu\nu}(x^a)$. Here Greek letters $\alpha, \beta, ...$ label spacetime indices that range over 0, 1, 2, 3, and Latin letters $a, b, c, ...$ label spatial indicies that range over 1, 2, 3.

\par The Lagrangian for a generally coupled scalar field is \cite{BD} 

\begin{equation}
L = 1/2 (-g)^{1/2} [ g^{\mu \nu} \Phi,_{\mu} \Phi,_{\nu} - (M^{2} + \xi R)\Phi^{2}], 
\end{equation}

\noindent where $M$ is the mass of the scalar particle, $R$ is the scalar curvature, and $\xi$ is the coupling ($\xi=0$ for minimal coupling and $\xi=1/6$ for conformal coupling.)\cite{BD} Varying the action $ S = \int L \ d^{3}x$ gives the Klein-Gordon field equation,

\begin{equation}
|g|^{-1/2}\left(|g|^{1/2}g^{\mu \nu} \Phi,_{\nu}\right),_{\mu} - \xi R\Phi=M^2\Phi.
\end{equation}

\noindent In the massless case with conformal coupling, this field equation is conformally invariant under a conformal transformation of the metric and field; in this case the inner product respecting the stress tensor for the field is also conformally invariant. This led Ishibashi and Hosoya \cite{IH} to state, in the context of wave regularity, that ``the calculation is as simple as that in the static case when singularities in conformally static space-times are probed with conformally coupled scalar fields.''

The conformally static metric has the form
\begin{equation}
ds^2 = g_{\mu\nu} dx^{\mu} dx^{\nu}
     = C^{2}(\eta) \gamma_{\mu \nu}(x^{c}) dx^{\mu} dx^{\nu}
     =  C^{2}(\eta)(\gamma_{\eta \eta} d\eta^{2} + \gamma_{ab}dx^{a} dx^{b}),
\end{equation} 

\noindent where $a,b,c = 1, 2, 3$. Then as shown by Kandrup\cite{K}, mode solutions $\chi, \zeta$ of a wave equation on the static portion of the metric (i.e., without the conformal factor $C^2(\eta)$) on a Hilbert space $\mathcal{L}^{2}(\Sigma)$ have the inner product

\begin{equation}
(\chi, \zeta) = \int d^{3} x (\gamma)^{1/2} (-\gamma_{\eta \eta})^{- 1/2} \chi(x^{a}) \zeta(x^{b}),
\end{equation} 
where $\gamma$ is the determinant of the spatial portion of the metric. 

\par At this point we consider the radial portion alone, change variables and write the radial equation in one-dimensional Schr\"odinger form, $H u(x) = E u(x)$,  where the operator $H = -d^2/dx^2 + V(x)$ and $E$ is a constant, with the singularity at $x=0$. The inner product here is simply $\int dx |u(x)|^2$ and the Hilbert space is $\mathcal{L}^2(0,\infty)$. One can now simply apply the limit point - limit circle criterion as easily as in the static case in order to determine the quantum singularity structure.

\section{A class of conformally static spacetimes}

We now specialize to a class of conformally static, spherically symmetric spacetimes with metrics of the form

\begin{equation}
ds^2 = a^2(t)\left(-f^2(r) dt^2 + \frac{dr^2}{f^2(r)} + S^2(r) d\Omega^2\right)
\end{equation}
where $d\Omega^2 = d\theta^2 + \sin^2\theta d\phi^2$, with the time-dependent conformal factor $a^2(t)$.  We study solutions of the generally coupled, massless Klein-Gordon equation of Eq. (7).  The curvature scalar is

\begin{eqnarray}
R = \frac{2}{a^3f^2S^2}[ - af^2  +   2aSS'' f^2 - 3S^2 \ddot a + aS'^2 f^4  \nonumber \\
                                         \ \ \ \ \ \ \ \ \           + 4af^3 f' S S' + S^2 a f^2 f'^2 + S^2 a f^3 f'' ]
\end{eqnarray}
where overdot $\equiv d/dt$ and prime $\equiv d/dr$. The Klein-Gordon equation separates into $\Phi \sim T(t) F(r) Y_{\ell m}(\theta, \phi)$, where the $Y_{\ell m}$ are spherical harmonics. The equation for $T(t)$ then becomes
\begin{equation}
\ddot T + 2\left(\frac{\dot a}{a}\right) \dot T + \left(6\ \xi \ \frac{\ddot a}{a} + q \right) T = 0,
\end{equation}
while the radial equation for $F(r)$ becomes

\begin{eqnarray}
 F'' &+&2\left( \frac{f'}{f} + \frac{S'}{S}  \right)F'    +  \left(\frac{q}{f^4} - \frac{\ell(\ell + 1}{f^2 S^2}\right) F  \nonumber\\ 
  & + & 2 \xi \left(- \frac{1}{f^2 S^2} + \frac{2S''}{S} + \frac{S'^2}{S^2} + \frac{f''}{f}  + \frac{f'^2}{f^2} + 4 \frac{f' S'}{f S}\right)F = 0,
\end{eqnarray}
where $q$ is a separation constant.

The first step in testing for quantum singularities is to convert the radial equation into a one-dimensional Schr\"{o}dinger equation

\begin{equation}
\frac{d^2u}{dx^2} + G(x) u = 0
\end{equation}
with normalization 

\begin{equation}
\int dr d\theta d\phi \ F^* F \sqrt{g_3/(- g_{00})} = 4\pi a^2 \int dx \ u^*(x) u(x) = 4\pi a^2.
\end{equation}
The conversion from $r$ to $x$ and $F(r)$ to $u(x)$ is carried out with  $F(r) = u(x)/S(r)$ and
\begin{equation}
dx/dr = 1/f^2(r)  \ \ \ \ (i. e., x = \int^r dr/f^2).
\end{equation}
We are then able to write $G(x) = E - V(x)$, where $E$ is a separation constant and the potential is  
\begin{eqnarray}
V(x) &=& \frac{f^2}{S} \frac{d}{dr}\left(f^2 \frac{dS}{dr} \right) + \frac{\ell (\ell + 1)}{S^2}f^2 
-  2\xi f^4 \Big[ \ 2\frac{d^2S/dr^2}{S}  + \frac{(dS/dr)^2}{S^2} \nonumber\\  
&+& \frac{(d^2f / dr^2 )}{f} + \frac{(df / dr)^2 }{f^2} + 4 \ \frac{(df/dr)( dS/dr)}{fS} - \frac{1}{f^2S^2}\Big].
\end{eqnarray}
We can then explore $V(x)$ near the singularity to see if a spacetime of this form is limit point  (LP) or limit circle (LC), using Theorem X.10 of Reed and Simon \cite{RS}.

To do so, let the singularity be located at $r = r_0$, and write an arbitrary radius in the form $r = r_0 + \epsilon$. Assume that both $f$ and $S$ can be represented to lowest order by a power law  in $\epsilon$ near the singularity, namely $f = C_0\epsilon^n$ and $S = C_1\epsilon^m$, so we can write $\epsilon$ in terms of $x$ by

\begin{eqnarray}
\epsilon &=& [C_0^2(1 - 2n)]^{1/(1 - 2n)} x^{1/(1 - 2n)} \ \ \ \  (n < 1/2) \\ \nonumber\\
\epsilon &=& e^{C_0^2  x} \ \ \  \ \ \ \ \ \ \ \ \ \ \ \ \ \ \ \ \ \ \ \ \ \ \ \ \ \ \ \ \ \ \ (n = 1/2)
\end{eqnarray}
where we confine ourselves to $n \le 1/2$ so that $\epsilon$ is real and so that the singularities occur at $\epsilon = 0$. For $n < 1/2$, the singularities are at $\epsilon = 0$ and $x = 0$;  for $n = 1/2$, the singularities are at $\epsilon = 0$ and $x = - \infty$. Substituting these into $V(x)$ gives, to lowest order,

\begin{eqnarray}
&V(x)& =  \frac{m(2n + m - 1) - 2 \xi [ m(3m - 2) + n(2n - 1) + 4nm]}{(1 - 2n)^2 x^2}  \nonumber\\
& + & \frac{[\ell(\ell + 1) + 2\xi ] \ C_0^2}{C_1^2}[C_0^2 (1 - 2n) x ]^{2(n - m)/(1 - 2n)}. \ \ (n < 1/2) \nonumber\\ \\\
&V(x)& = m^2C_0^4 + \frac{\ell (\ell + 1) C_0^2}{C_1^2} e^{(1 - 2m) C_0^2 x} \nonumber\\
&-& 2\xi C_0^4\left[3m^2 - \frac{1}{C_0^2C_1^2} e^{(1 - 2m) C_0^2 x}  \right]. \ \ \ \  \ \ \ \ \ \ \ \ \ \ \ \ \ \ \ (n = 1/2)
\end{eqnarray}

Consider now the $n < 1/2$ case. The singularity is at $x = 0$, and the potential $V(x)$ is in the LP case if $V(x) \ge 3/4x^2$ near the singularity; otherwise it is LC. First, suppose that $n + m < 1$, so that the first term in $V(x)$ dominates. We can therefore characterize the LC and LP regions in an $(m, n)$ parameter plane, for any given value of the coupling constant $\xi$.  If $\xi = 0$, i.e., the coupling is minimal, then the three LC regions correspond to the following, each with $n < 1/2$: (a) $n - 1/2 < m < 0$, (b) $0 < m < - 3n + 3/2 <  3/4$, and $3/4 < m < - n + 1$. The two LP regions correspond to (a) $m < n - 1/2 < 0$ and (b) $- 3n + 3/2 < m < - n + 1$, both with $n < 1/2$.  If instead $\xi = 1/6$, i.e., the coupling is conformal, then  $V(x)$ is in the LP case if $m \le (11/2)n - 9/4$, and in the LC case if $ - n + 1 > m > (11/2)n - 9/4$, again restricted to $n < 1/2$. The various regions can easily be worked out for other choices of coupling constant if desired. 

Now suppose that $n + m > 1$, so the second term in $V(x)$ dominates. Then $V(x)$ is in the LP case if $\ell(\ell + 1) + 2\xi > 0$, and in the LC case if  $\ell(\ell + 1) + 2\xi < 0$. If $\ell(\ell + 1) + 2\xi = 0$, the second term vanishes, so the first term dominates regardless of $n + m$.  

In the $n = 1/2$ case the singularity is at $x = - \infty$, and it is null. Therefore the $n = 1/2$ spacetime is globally hyperbolic, the spatial Klein-Gordon operator is essentially self-adjoint, and the potential is necessarily in the LP case \cite{HM}. The use of quantum particles heals the classical singularity in this case, for all values of the other parameters.    

\section{Special cases}

As special cases we consider three important dynamical and spatially inhomogeneous exact solutions to Einstein's equations. All involve massless scalar fields coupled to gravity. These spacetimes can be taken as simple models of the collapse of matter to form or grow black holes or naked singularities, a process that normally requires numerical studies to analyze the complicated differential equations that arise. The Einstein-scalar field spacetime models in this section were constructed by Roberts \cite{R} in 1989, Husain, Martinez, and N\'u\~nez (HMN) \cite{HMN} in 1994, and Fonarev  \cite{F} in 1995. The energy-momentum tensor for the minimally-coupled scalar field has in each case the form

\begin{equation}
T_{\mu \nu} = \left( \phi,_{\mu}\phi,_{\nu} - \frac{1}{2}g_{\mu \nu} \phi,_{\alpha}\phi^{,\alpha}\right) - g_{\mu \nu}U
\end{equation}
which yields the Einstein equation
\begin{equation}
R_{\mu \nu} = 8 \pi\left( \phi,_{\mu}\phi,_{\nu} + g_{\mu \nu}U\right).
\end{equation}
Here $\phi$ is the scalar field and $U$ is its potential. The potential is zero for both the Roberts and HMN spacetimes, and $U= U_0 e^{- \sqrt{8 \pi} \lambda \phi}$ for Fonarev, where $U_0$ and $\lambda$ are constants. Now we consider each metric in turn;  all are conformally static and spherically symmetric.
\\[0.3cm]
\noindent (1) The Roberts spacetime \cite{R} has metric coefficients $a(t) = e^t, f(r) = 1$, and  
\begin{displaymath}
S^2(r) = \frac{1}{4}\left[1 + p - (1 - p)e^{-2r}\right](e^{2r} - 1) \ \ \  \mathrm{with} \ \ \  0 < p < 1,
\end{displaymath}
The spacetime is self-similar, with a massless scalar field that models collapse to a timelike naked singularity. It has a timelike scalar curvature singularity at $r = 0$ and a global structure that is identical to that of the negative-mass Schwarzschild solution. The existence of a scalar curvature singularity requires not only that some scalar in the curvature diverge, but also that either null or timelike geodesics (or both) reach the singularity with a finite affine parameter $\lambda$. In fact, the curvature scalar $R$ diverges as $r \rightarrow 0$ for Roberts, so the first test is met. 

It is straightforward to show that inward-directed radial null geodesics in the general class of spacetimes we are considering possess the two first integrals of motion $dr/d\lambda = - K/a^2(t)$ and $dt/d\lambda = K/a^2(t)f^2(r)$, where $K$ is a positive constant. Therefore the affine length of a null geodesic starting at radius $r_0$ at time $t_0$ and ending at smaller radius $r$ is

\begin{equation}
\lambda = \frac{1}{K} \int_r^{r_0} dr \ a^2\left(t_0 + \int_r^{r_0} dr/f^2(r)\right)
\end{equation}
where the notation means that the quantity $t_0 + \int_r^{r_0} dr/f^2(r)$ is to be substituted for $t$ in $a(t)$. For the Roberts spacetime, with $f(r) = 1$ and $a(t) = e^t$, it follows that

\begin{equation}
\lambda = \frac{1}{K} \int_r^{r_0} dr \ e^{2\left(t_0 + \int_r^{r_0} dr \right)} = \frac{1}{2K}e^{t_0}\left[e^{2(r_0 - r)} - 1\right],
\end{equation} 
which is finite as $r \rightarrow 0$. Therefore there is a classical scalar curvature singularity at $r = 0$. Is $r = 0$ also quantum mechanically singular?

Substituting the quantities $f(r)$ and $S(r)$ into the general potential $V(x)$, it is straightforward to show, using Theorem 2 of Section 2, that there exists a positive \emph{constant} ($M$) such that all three criteria listed in Theorem 2 criteria are satisfied, for any finite value of $\ell$ and $\xi$. Therefore the Roberts spacetime is complete at infinity.
 
In terms of our general form, the Roberts parameters describing $f(r)$ and $S(r)$ near $r = 0$ are $C_0 = 1$, $n = 0$, $C_1 = \sqrt{p}$, and $m = 1/2$. Therefore as $x \rightarrow 0$,
\begin{equation}
V(x) \sim \frac{-1 + 2\xi}{4x^2} \ \ \ \ (\xi \ne 1/2)
\end{equation} 
\begin{equation}
V(x) \sim \frac{\ell^2 + \ell + 1}{px} \ \ \ \ (\xi = 1/2)
\end{equation} 
It follows from Theorem 3 of Section 2 that the Roberts spacetime is LC (quantum mechanically singular) unless $\xi \ge 2$, in which case it is LP  (quantum mechanically nonsingular). In particular, the spacetime is LC for both the minimally coupled and conformally coupled cases. Quantum mechanics fails to heal the classical singularity in these cases.

The time equation for the Roberts spacetime is

\begin{equation}
\ddot T + 2 \dot T + \left(6 \xi + q \right) T = 0,
\end{equation}
which is a damped harmonic oscillator equation with solutions
\begin{equation} 
T(t) = e^{-t} (A e^{\sqrt{1 - 6\xi - q}t}  + B e^{- \sqrt{1 - 6\xi - q}t} )
\end{equation}
for the overdamped ($6\xi -q < 1$) and underdamped ($6\xi -q > 1$) cases, and
\begin{equation} 
T(t) = A e^{-t} (1 + Bt )
\end{equation}
for the critically damped case ($6\xi -q = 1$), where $A$ and $B$ are constants. 

If we had chosen the conformal factor to be $a(t) = \mathrm{constant}$, corresponding to a class of \emph{static} spacetimes, the time equation would be simply the harmonic oscillator equation. It is interesting that the Roberts conformally static spacetime converts the time portion of the wave equation from a simple harmonic oscillator for the corresponding static spacetime, to that of a damped harmonic oscillator. In any case the time solutions do not influence whether or not a conformally static spacetime is quantum mechanically singular.
\\[0.3cm]
(2) Husain, Martinez, and N\'u\~nez (HMN) spacetimes \cite{HMN}, have metric coefficients $a(t) = \sqrt{- t + b}$, $f(r) = \left(1 - 2/r\right)^{\alpha/2}$, and $S^2(r) = r^2\left(1 - 2/r\right)^{1 - \alpha}$ (where $\alpha = \pm\sqrt{3}/2$ only.) The HMN spacetimes were constructed to model scalar field collapse. They have the conformal Killing vector $v = \partial/\partial t$ as well as the three Killing vectors associated with spherical symmetry. Since the spacetime is not asymptotically flat (but \emph{is} asymptotically conformally flat), the authors interpret it to be an inhomogeneous scalar field cosmology. We are considering only the nonstatic solutions, which have scalar curvature singularities at $r = 2$ and $t = - b/a$ (where $a = \pm 1$) for both $\alpha = \sqrt{3}/2$ and $\alpha = - \sqrt{3}/2$. Without loss of generality $b$ can be set equal to zero, so then the time coordinate ranges over $0 \le  t < \infty$ for $a = 1$ and $- \infty < t \le 0$ for $a = -1$. These correspond to expanding and collapsing scalar fields. HMN consider only the collapsing case, in which case the spacetime has a timelike scalar curvature singularity at $r = 2$ (that is, at the areal radius $R = rS = 0$) and a spacelike singularity at $t = 0$. The coordinate $r$ ranges over $2 \le r < \infty$.  Using $f(r)$ and $a(t)$ for the HMN metric in Eq. 24, one can show that radial null geodesics reach the $r = 2$ timelike singularity with finite affine length, so $r = 2$ fulfills all necessary criteria for being a classical scalar curvature singularity.

Substituting the quantities $f(r)$ and $S(r)$ into the general potential $V(x)$, it is straightforward to show, using Theorem 2 of Section 2, that there exists a positive \emph{constant} ($M$) such that all three criteria listed in Theorem 2 criteria are satisfied, for any finite value of $\ell$ and $\xi$. Therefore the HMN spacetimes are also complete at infinity.

In terms of our general form, the HMN parameters describing $f(r)$ and $S(r)$ near the timelike singularity at $r = 2$ are $C_0 = (2)^{- \alpha/2}$, $n = \alpha/2$, $C_1 = (2)^{(\alpha + 1)/2}$, and $m = (1 - \alpha)/2$, where in each case $\alpha = \pm\sqrt{3}/2$ only.  In these cases the potential becomes
\begin{equation}
V(x) = - \frac{1}{4x^2} + \frac{\xi}{2}\left( \frac{1 \pm \sqrt{3}/2}{1 \mp \sqrt{3}/2} \right)\frac{1}{x^2}+ \left[\ell(\ell + 1) + 2 \xi \right](1 \mp \frac{\sqrt{3}}{2} )x^{\frac{\pm\sqrt{3} - 1}{1 \mp \sqrt{3}/2}}.
\end{equation}
Regardless of the choice of signs, the $1/x^2$ terms dominate, so the HMN spacetimes are LP if $\xi \ge 2[(1 + \sqrt{3}/2)/(1 - \sqrt{3}/2)]$, and are otherwise LC. So in particular they are LC, and therefore quantum mechanically singular, for both the minimally coupled and conformally coupled cases.   

The time equation for the HMN metric is Bessel's equation
\begin{equation}
\ddot T - \frac{1}{(-t + b)} \dot T + \left[\frac{- 3\xi}{2}\frac{1}{(-t + b)^2} + q\right]T = 0.
\end{equation}
\\[0.3cm]
\noindent (3) Fonarev spacetimes \cite{F, Maeda} have metric coefficients $a(t) = |t/t_0|^{2/(\lambda^2 - 2)}$, $f(r) = (1 - 2w/r)^{\alpha/2}$, and  $S^2(r) = r^2(1 - 2w/r)^{1 - \alpha}$ (where $0 < \alpha^2 \le 3/4$, but $\alpha^2 \ne 1/2$.)  Our index $n$ is less than 1/2 in all the Fonarev solutions. The parameter $\lambda$, a so-called \emph{steepness parameter}, satisfies $0 < \lambda^2 \le 6$ with $\lambda^2 \ne 2$ for a non-negative potential; $\lambda$ is related to $\alpha$ by $\alpha = \lambda/\sqrt{\lambda^2 + 2}$. If $\lambda^2 = 6$, the solution reduces to HMN.  The Fonarev spacetimes are a generalization of the HMN spacetimes, to include a scalar field potential, as in Eq. [22].  Massless scalar fields with exponential potentials appear in supergravity theories \cite{Townsend} and in dimensionally reduced effective four-dimensional theories \cite{GL, EG}, as explained by Maeda \cite{Maeda}  in his paper on the physical interpretation of the Fonarev solution. 

If the parameter $w = 0$, the Fonarev solution reduces to the flat Friedmann-Robertson-Walker cosmology. Here we consider $w \ne 0$, which becomes the FRW spacetime as $r \rightarrow \infty$. The coordinate $r$ ranges $2w < r < \infty$ for $w > 0$ and $0 < r < \infty$ for $w < 0$. For $w \ne 0$, there are central scalar curvature singularities at $r = 0$ and at $r = 2w$. For $0 < \lambda^2 < 2$, there is a null curvature singularity as $t \rightarrow \pm \infty$, while for $2 < \lambda^2 \le 6$ there as a spacelike curvature singularity at $t = 0$. Thus the $t$ - coordinate ranges are $\infty < t < 0$ and $0 < t < \infty$, with big bang and big crunch singularities. We will study only the locally-naked timelike central singularities.\footnote{According to Hayward \cite{Hayward}, if in the collapsing case the solution has a trapping horizon, then it represents a dynamical black hole. However, as Maeda \cite{Maeda} notes, here the singularity inside the trapping horizon is a big crunch spacelike singularity, and not the timelike singularity which can then be considered naked.} Inward-directed null geodesics reach the timelike singularity at $r = 2w$ in finite affine parameter according to Eq. 24, so $r = 2w$ fulfills necessary requirements to be a classical scalar curvature singularity.
\\[0.3cm]
In terms of our general form, the Fonarev parameters describing $f(r)$ and $S(r)$ near the timelike singularity at $r = 2w$ are $C_0 = (2w)^{- \alpha/2}$, $n = \alpha/2$, $C_1 = (2)^{(\alpha + 1)/2}(w)^{(\alpha - 1)/2}$, and $m = (1 - \alpha)/2$, where in each case $0 < \alpha^2 \le 3/4$, but $\alpha^2 \ne 1/2$. Using these parameter values, it is then straightforward to show from Eq.[20] that the potential $V(x)$ for these spacetimes is in the LC case if
\begin{equation}
\xi  <  2\left(\frac{1 - \alpha}{1 + \alpha} \right) 
\end{equation}
and in the LP case if
\begin{equation}
\xi \ge 2\left(\frac{1 - \alpha}{1 + \alpha} \right). 
\end{equation}
In particular, Fonarev spacetimes are LC for both the minimally and conformally coupled cases.

For the Fonarev spacetime, the time equation is again Bessel's equation,
\begin{equation}
\ddot T + \frac{4}{(\lambda^2 - 2)t} \dot T + \left[\frac{12 \xi}{t^2}\frac{(4 - \lambda^2)}{(\lambda^2 - 2)^2} + q\right]T = 0.
\end{equation}
\\[0.3cm]

\section{Conclusions}

We have confirmed that the Horowitz and Marolf definition of quantum singularities for static spacetimes can be extended to the case of conformally static spacetimes. We then tested the formalism for a class of conformally static, spherically symmetric spacetimes, a class that includes the special cases of spacetimes studied by Roberts, by Fonarev, and by Husain, Martinez, and N\'u\~nez (HMN).  We used as quantum fields the solutions of the generally coupled, massless Klein-Gordon equation, and Weyl's limit point - limit circle criteria for judging the existence of quantum singularities. This required that we write the radial part of the Klein-Gordon equation in the form of a one-dimensional Schr\"odinger equation, and evaluate the behavior of the associated potential energy in the vicinity of the singularity.  

In this way we found that the Roberts spacetimes are limit circle (i. e., quantum mechanically singular) for both the minimally coupled (coupling parameter $\xi = 0$) and conformally coupled ($\xi = 1/6$) cases, and in general if the coupling parameter $\xi < 2$; they are limit point (quantum mechanically nonsingular) if $\xi \ge 2$. The HMN spacetimes are limit circle if $\xi < 2[(1 + \sqrt{3}/2)/(1 - \sqrt{3}/2)]$, and so are also limit circle for both the minimally coupled and conformally coupled cases.  
 Similarly, Fonarev spacetimes are limit circle if $\xi < 2(1  - \alpha)/(1 + \alpha)$, where the parameter $\alpha$ in these spacetimes is constrained by $0 < \alpha^2 \le 3/4$, but $\alpha^2 \ne 1/2$. In particular, all Fonarev spacetimes are limit circle for both minimally coupled and conformally coupled cases. 
 
As in the studies of other metrics up to now, we have found that the use of the Horowitz and Marolf procedure is successful in healing classical singularities for some parameter values in the metrics, but not for others. One goal for the future is to understand more comprehensively and more deeply which singularities can be healed in this way, and which cannot, and why.

 
\bibliography{HK2013references}

\end{document}